\documentclass[pdflatex,sn-nature]{sn-jnl}

\usepackage{graphicx}
\usepackage{amsmath,amssymb,amsfonts}
\usepackage{xcolor}

\usepackage{bm}
\hypersetup{hypertexnames=false,hidelinks}

\begin{document}

\title{Attosecond circular-dichroism spectroscopy of hole ring currents}

\author[1,2]{\fnm{Guangru} \sur{Bai}}

\author[1,2]{\fnm{Zhihui} \sur{Lyu}}

\author[1,2]{\fnm{Jinlei} \sur{Liu}}

\author*[1,2]{\fnm{Jing} \sur{Zhao}}\email{jzhao@nudt.edu.cn}

\author*[1,2]{\fnm{Zengxiu} \sur{Zhao}}\email{zhaozengxiu@nudt.edu.cn}

\affil[1]{College of Science, National University of Defense Technology, Changsha 410073, China}
\affil[2]{Hunan Key Laboratory of Extreme Matter and Applications, Changsha 410073, China}

\abstract{\unboldmath
 Ultrafast ionization of atoms by circularly polarized few-cycle laser pulses generates a hole ring currents, offering a route for ultrafast manipulation of magnetism. Here we explore the subcycle formation of these currents whose circulation direction is determined by the driving-field helicity, whereas their magnitude is governed by the quantum coherence of the residual ion. We show that the correlated ion-photoelectron dynamics can be probed with attosecond transient-absorption circular dichroism. Using the Wigner-Eckart theorem, we derive a linear relation between state-resolved dichroic absorption and the orbital and spin angular-momentum projections of the hole, thereby extending the Thole-Carra magnetic CD sum rules from static X-ray spectroscopy to attosecond electron dynamics. The modulation depth of the dichroic signal provides a direct optical measure of the coherence injected by ionization, and its dependence on the pump-pulse duration reveals a competition between coherence build-up and ionization-induced dephasing. These findings demonstrate the potential of attosecond circular dichroism for probing ionic coherence and highlight the role of quantum coherence in ultrafast angular momentum exchange between light and matter.
}

\maketitle

\section*{Introduction}
The manipulation of magnetism by ultrashort laser pulses has become a fundamentally challenging topic with high-impact applications \cite{Beaurepaire1996,Kirilyuk2010}. As magnetism originates primarily from electronic orbital and spin angular momentum, circularly polarized light carrying a definite optical angular momentum can create a ring current that initiates the subsequent charge, spin and lattice dynamics toward light-driven (de)magnetization \cite{Siegrist2019}. Probing and understanding the angular-momentum transfer between an atom and light thus provides important insight into the quantum excited non-equilibrium state at the earliest stage \cite{vazXrayMagneticCircular2025}.

In spite of the lack of a first-order interaction of light with spin, single-photon ionization by circularly polarized light can produce spin-polarized electrons because the spin-orbit coupling modifies the orbital wave function (Fano effect \cite{Fano69}) or the density of states. On the other hand, a hole ring current is created in the residual ion with a circulation direction depending on the sign of the magnetic quantum number $m$. When subjected to intense laser fields, tunnelling ionization of atoms by absorbing an indefinite number of photons creates a swift ring current within each optical cycle \cite{hartungElectronSpinPolarization2016a,eckartUltrafastPreparationDetection2018b,trabertSpinAngularMomentum2018,herathStrongfieldIonizationRate2012a}. It has been shown theoretically and confirmed experimentally that strong-field tunnel ionization of noble-gas atoms preferentially removes counter-rotating electrons \cite{barthNonadiabaticTunnelingCircularly2011,barthNonadiabaticTunnelingCircularly2013b,barthComparisonTheoryExperiment2013}, so the driving-field helicity controls both the hole ring current and the resulting photoelectron spin polarization \cite{Carlstrom2025,zhangSpinPolarizationStrongfield2025,liuEnergyMomentumresolvedPhotoelectron2018}. Interestingly, the strong-field-created ion and the photoelectron are entangled \cite{kollEntanglementElectronicCoherence2026,Laurell2025}, offering new possibilities for quantum manipulation at the single-atom level or in bulk ferromagnetic materials. However, resolving the subcycle ring-current dynamics remains a challenging task \cite{eckartUltrafastPreparationDetection2018b,hanAttosecondCirculardichroismChronoscopy2022a,smirnovaHighHarmonicInterferometry2009,cireasaProbingMolecularChirality2015,baykushevaBicircularHighHarmonic2016,neufeldBackgroundFreeMeasurementRing2019}, and calls for circular-dichroism (CD) probing with attosecond resolution.

When the photoelectron is not observed, tracing over it leaves the vacancy in a partially coherent reduced state \cite{goulielmakisRealtimeObservationValence2010}. The hole ring current is thus directly related to the degree of quantum coherence of the residual ion, which is crucial to coherent magnetism and chirality \cite{vazXrayMagneticCircular2025,beaulieuAttosecondresolvedPhotoionization2017,Siegrist2019}. Solely measuring the density of the electronic wave packet cannot distinguish ring currents of opposite direction. Previous detection schemes based on CD generally fall into two categories. The first is based on electron spectroscopy, in which a circularly polarized pump prepares the ring current and a secondary circularly polarized pulse photoionizes the system \cite{eckartUltrafastPreparationDetection2018b,hanAttosecondCirculardichroismChronoscopy2022a}. However, such measurements are intrinsically destructive and of limited time resolution. The second is based on high-harmonic spectroscopy \cite{smirnovaHighHarmonicInterferometry2009,cireasaProbingMolecularChirality2015,baykushevaBicircularHighHarmonic2016}, in which a ring current breaks the reflection symmetry of the driven atom, so that the emitted harmonics acquire ellipticity. The ellipticity provides only an indirect, pulse-integrated measure of the ring current, and the mapping between ring-current magnitude and harmonic ellipticity is quadratic \cite{neufeldBackgroundFreeMeasurementRing2019}. Neither scheme therefore follows the ring-current dynamics on the subcycle time scale with state resolution as the currents are created and evolve through coherent motion among spin-orbit-split states \cite{goulielmakisRealtimeObservationValence2010,WangLi2024}. In particular, the coherent ring current presumed to persist after the pulse has not been resolved, and its absence in the measurement has remained unexplained \cite{eckartUltrafastPreparationDetection2018b}.

Here we explore the build-up of a coherent hole ring current in Kr by a few-cycle optical pulse and propose its real-time tracking by the circular dichroism of attosecond transient-absorption spectroscopy (ATAS) \cite{Siegrist2019,drescherAttosecondOpticalOrientation2025a}. The rapid development of circularly polarized attosecond pulse generation \cite{kfirGenerationBrightPhasematched2015a,hanAttosecondMetrologyCircular2023a} and transient-absorption spectroscopy~\cite{geneauxSpinDynamicsMetallic2024a,Siegrist2019,willemsProbingUltrafastSpin2015a,WangLi2024} enables probing attosecond dynamics as the spatial or temporal symmetry is broken. X-ray magnetic circular dichroism has been shown to be capable of quantifying static spin and orbital moments \cite{vazXrayMagneticCircular2025} thanks to the powerful sum rules \cite{Thole1992,Carra1993}. Whether these sum rules are applicable to a coherently excited state with a time-varying current has lacked solid proof. Combining theory and numerical simulation, we confirm that the attosecond CD absorption signal depends linearly on the projections of the orbital and spin angular momenta, thus lifting the conceptual obstacles and generalizing the X-ray magnetic circular dichroism sum rules to attosecond dynamics. We find that the coherence induced by strong-field ionization is controlled by the pulse-duration-dependent ionization injection and by the Raman coupling of the hole with the pump pulse, both of which play key roles in the ring-current formation. Because the vacancy coherence carries information on the ion-photoelectron entanglement, the present scheme is also valuable for probing entanglement dynamics during strong-field ionization.

\section*{Results}

\begin{figure}[!htbp]
\centering
\includegraphics[width=0.98\textwidth]{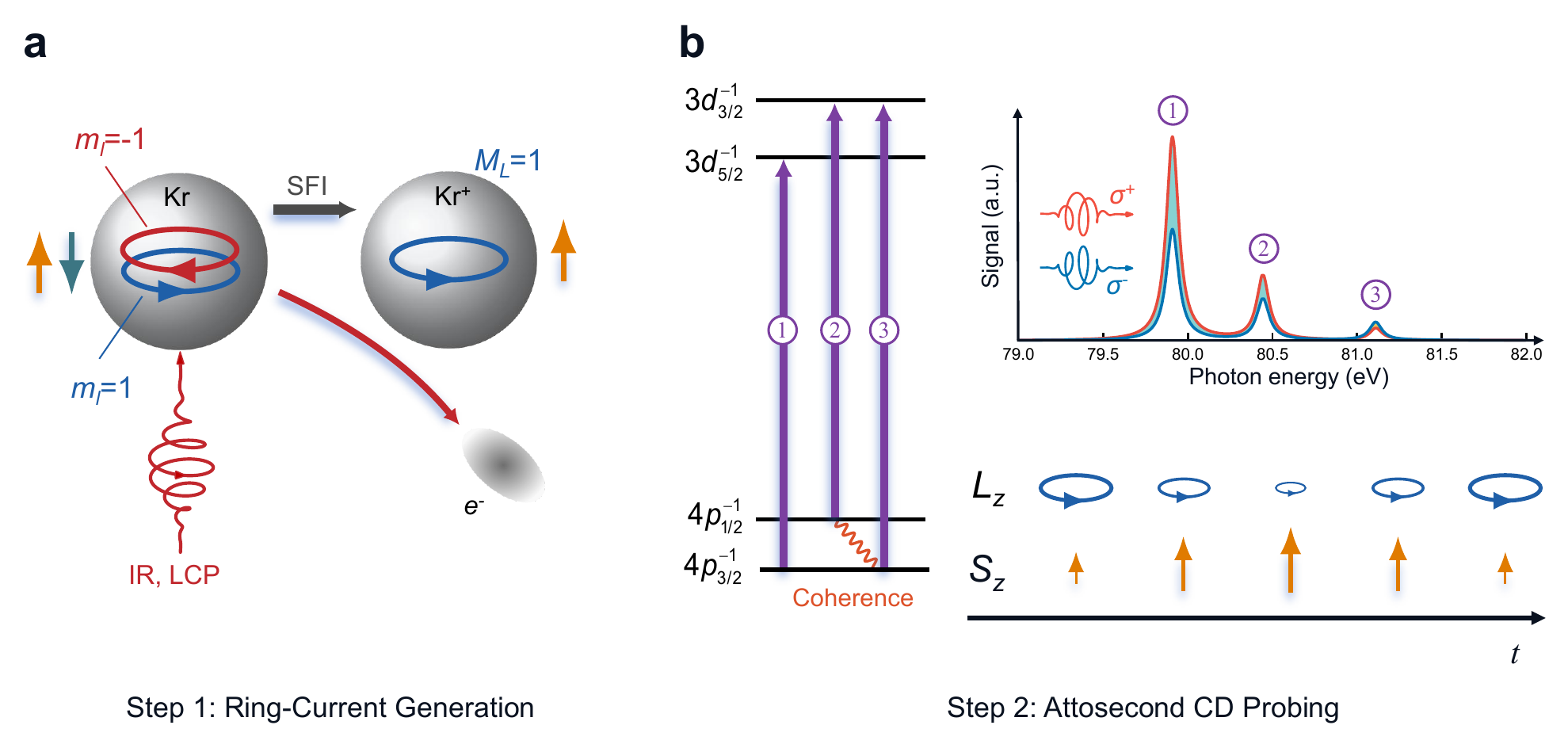}
\caption{Pump-probe scheme for generating and probing valence-vacancy ring currents and spin dynamics in krypton ions.
	(a) A left-circularly polarized (LCP) femtosecond pump pulse induces strong-field ionization of krypton, creating an entangled hole ring-current state and photoelectron wave packet within each optical cycle.
	(b) Through spin-orbit coupling, the initially prepared hole ring-current state evolves into coupled orbital and spin dynamics involving the $4p^{-1}_{3/2}$ and $4p^{-1}_{1/2}$ manifolds. Attosecond circularly polarized XUV probe pulses with opposite helicities interrogate the system via the $4p^{-1}\rightarrow 3d^{-1}$ core-level transitions through three paths:
	\textcircled{1} $4p_{3/2}^{-1}\rightarrow 3d_{5/2}^{-1}$,
	\textcircled{2} $4p_{1/2}^{-1}\rightarrow 3d_{3/2}^{-1}$, and
	\textcircled{3} $4p_{3/2}^{-1}\rightarrow 3d_{3/2}^{-1}$.
	By scanning the pump-probe delay and comparing absorption spectra obtained with left- and right-circularly polarized XUV probes, the transient CD signal maps the amplitude, handedness, and temporal evolution of the ring current and the accompanying spin dynamics. The modulation of the ring current reflects the coherence between the two spin-orbit-split states.}
\label{fig:1}
\end{figure}

\subsection*{Measuring scheme}
The all-optical pump-probe scheme for tracking electronic ring currents and their coupled orbital-spin dynamics is illustrated in Fig.~\ref{fig:1}. An intense few-cycle left-circularly polarized (LCP) pump pulse ($5\,\mathrm{fs}$, $2\times10^{14}\,\mathrm{W/cm^2}$) is applied to krypton atoms.
Owing to the $m_l$ selectivity of nonadiabatic tunneling ionization, the $4p$ electron with $m_l=-1$ is preferentially removed \cite{barthNonadiabaticTunnelingCircularly2011}. This prepares a valence vacancy in $\mathrm{Kr}^+$ with orbital-angular-momentum projection $M_L=1$, corresponding to a counterclockwise ring current. Spin-orbit (SO) coupling projects this orbital state onto the SO-split vacancy manifolds $4p_{3/2}^{-1}$ and $4p_{1/2}^{-1}$. For $M_J=3/2$ the vacancy lies entirely in $4p_{3/2}^{-1}$, producing a static ring current with fixed spin polarization, whereas for $M_J=1/2$ the vacancy forms a coherent superposition of $4p_{3/2}^{-1}$ and $4p_{1/2}^{-1}$ \cite{goulielmakisRealtimeObservationValence2010}. In the latter case, the two components are spin-orbit coupled, so that the orbital and spin dynamics cannot be separated. As illustrated in Fig.~\ref{fig:1}(b), the spin-orbit splitting ($\Delta_{\rm SO}=0.67$~eV) drives a periodic evolution of the ionic coherence, producing oscillations of the charge and spin densities, i.e., a time-dependent ring current with an antiphase spin modulation.

To access these coupled dynamics, we apply time-delayed attosecond XUV probe pulses ($150\,\mathrm{as}$, $80\,\mathrm{eV}$) with opposite helicities and record transient-absorption spectra. The resulting CD signal is delay-independent for the $M_J=3/2$ contribution but exhibits an oscillatory modulation for the coherent $M_J=1/2$ contribution. By analyzing the CD signal as a function of pump-probe delay, we reconstruct the instantaneous direction, magnitude, and temporal evolution of the ionic ring current together with the accompanying spin dynamics. In contrast to attosecond CD chronoscopy, which probes continuum-continuum CD via photoelectron spectroscopy \cite{hanAttosecondCirculardichroismChronoscopy2022a}, our scheme exploits resonant bound-bound absorption, and the CD signal is governed entirely by ionic-state observables and is insensitive to continuum phase distortions.

\subsection*{Theoretical model}
The ionic dynamics can be described theoretically by an ionization-coupling master equation for the density matrix $\rho^{+}$,
\begin{equation}
\dot{\rho}^{+}
=
-i\bigl[H(t),\rho^{+}\bigr]
+S_{\rm ion}(t)
+D_{\rm decay}(t),
\label{eq:one}
\end{equation}
where $S_{\rm ion}(t)$ describes ionization injection, with rates from Ref.~\cite{barthNonadiabaticTunnelingCircularly2011}, and $D_{\rm decay}(t)$ accounts phenomenologically for Auger decay and the associated dephasing of core-excited coherences. $H(t)$ contains the dipole couplings of the ion to both the pump and the attosecond probe pulses, with the latter delayed by $t_d$ relative to the pump. Atomic units are used throughout. Computational details are given in Methods and in the Supplementary Information.

\begin{figure}[!htbp]
\centering
\includegraphics[width=0.78\linewidth]{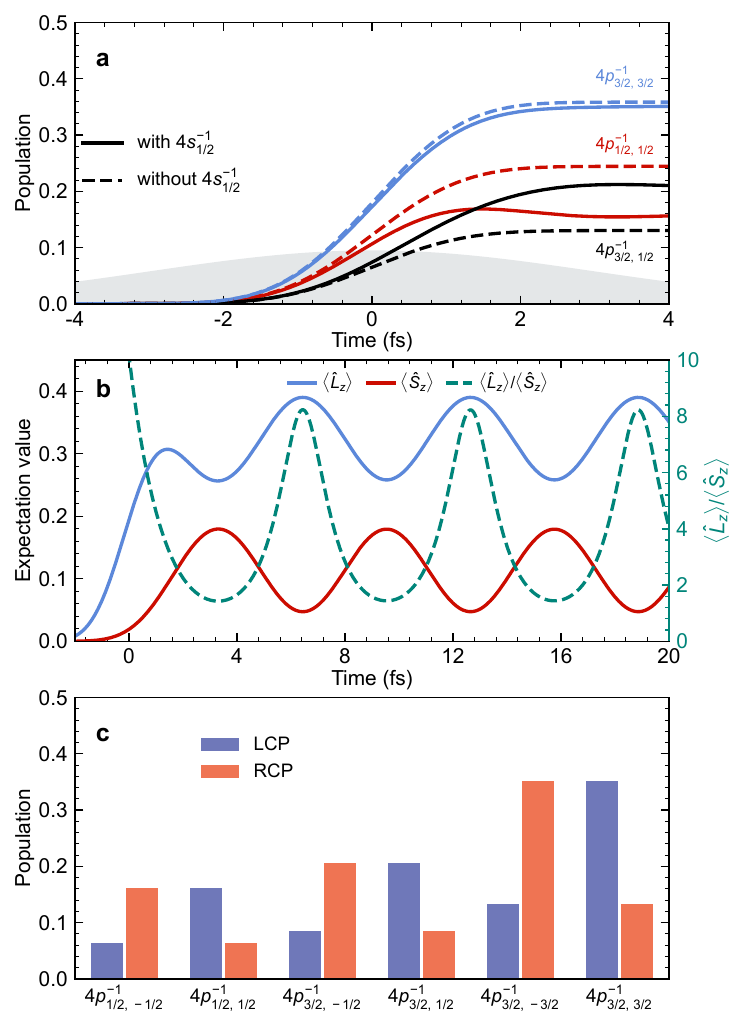}
\caption{
Strong-field preparation of population and angular momentum in the spin-orbit-split $4p^{-1}$ vacancy states of Kr$^+$.
(a) Time-dependent populations of the $M_J>0$ components in the $4p^{-1}$ manifold during the LCP pump pulse. Solid and dashed curves denote the results with and without the $4s_{1/2}^{-1}$ coupling, respectively.
(b) Time-dependent expectation values of the orbital angular momentum $\hat L_z$ and spin angular momentum $\hat S_z$ (left axis), together with their ratio $\langle\hat L_z\rangle/\langle\hat S_z\rangle$ (right axis).
(c) Final relative populations of the magnetic sublevels in the $4p^{-1}$ manifold after LCP and RCP pumping. Reversing the pump helicity exchanges the populations of opposite-sign magnetic sublevels.
}
\label{fig:2}
\end{figure}

In addition to including the instantaneous injection of $4p_{3/2}^{-1}$ and $4p_{1/2}^{-1}$ by strong-field ionization, our model further incorporates a Raman coupling between the $4p^{-1}$ vacancy manifold and the deeper $4s_{1/2}^{-1}$ ionic state, mediated by the strong infrared pump through sequential absorption and emission of pump photons \cite{WangLi2024}. This Raman pathway is essential to the scheme, since it actively transfers population between the spin-orbit-split substates inside the $M_J=1/2$ manifold and thereby enhances the ionic spin polarization that drives the CD signal. Figure~\ref{fig:2}(a) shows the resulting pump-driven population dynamics of the $4p^{-1}$ vacancy states. The Raman-induced transfer from $4p_{1/2}^{-1}$ to $4p_{3/2}^{-1}$ (solid vs.\ dashed curves) enhances the $M_S=+1/2$ contribution and the corresponding ionic spin polarization. The population with $M_J<0$, originating from ionization of the $m_l=+1$ electron, exhibits symmetric dynamics (see Supplementary Information). Although circularly polarized ionization is continuous in time and lacks the subcycle gating of linearly polarized fields, the same Raman mechanism operates here through the interplay of the SO coherence and the rotating IR field. Figure~\ref{fig:2}(b) shows the orbital and spin angular momenta, $\langle \hat L_z\rangle$ and $\langle \hat S_z\rangle$, together with their ratio $\langle \hat L_z\rangle/\langle \hat S_z\rangle$. During the initial stage of ionization, the ionic angular momentum is progressively built up. At later times, after ionization is completed, $\langle \hat L_z\rangle$ and $\langle \hat S_z\rangle$ exhibit pronounced out-of-phase oscillations, while their sum $\langle\hat J_z\rangle=\langle \hat L_z\rangle+\langle \hat S_z\rangle$ remains constant. This behavior reflects the spin-orbit-driven periodic redistribution of angular momentum between the orbital and spin degrees of freedom. The initially large $\langle \hat L_z\rangle/\langle \hat S_z\rangle$ reflects the preferential build-up of orbital angular momentum by circularly polarized ionization, further amplified by the small initial $\langle \hat S_z\rangle$. After the pump pulse, the final relative populations of the $4p^{-1}$ vacancy manifolds are shown in Fig.~\ref{fig:2}(c). By mirror symmetry under reversal of the pump helicity, LCP and RCP pumping yield identical populations for opposite $M_J$ at fixed $J$.

\begin{figure}[!htbp]
	\centering
	\includegraphics[width=0.95\linewidth]{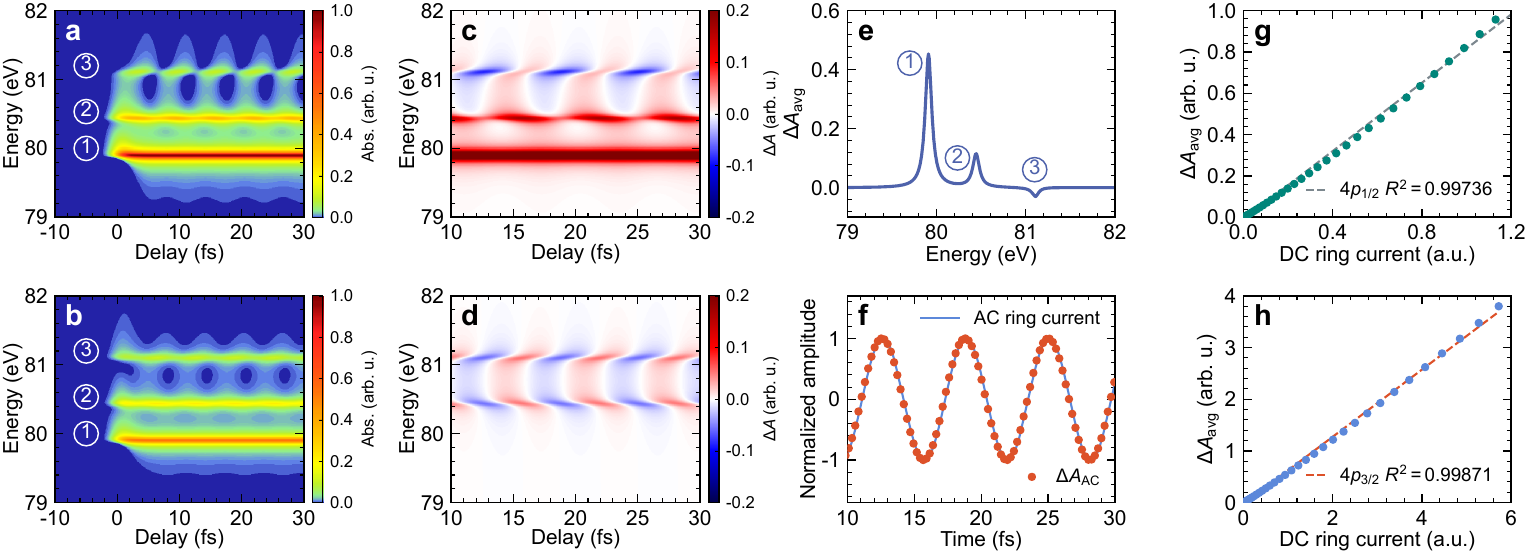}
	\caption{
		Attosecond transient-absorption spectra and circular-dichroic mapping of the vacancy ring current.
		(a,b) Delay-dependent attosecond transient-absorption spectra of Kr$^+$ probed by left- and right-circularly polarized attosecond XUV pulses, respectively.
		(c) Delay-dependent circular-dichroic absorption difference.
		(d) ac component of the CD absorption difference obtained by subtracting the delay-averaged component from the CD signal in (c).
		(e) Delay-averaged component of the CD absorption difference in (c), with three peaks at the resonant transition energies of the three paths.
		(f) Comparison between the energy-integrated ac component of the circular dichroism and the ac ring current as a function of pump-probe delay, with both curves independently normalized to unit peak amplitude for shape comparison.
		(g,h) State-resolved dc CD absorption differences versus the corresponding dc ring currents for the $4p_{1/2}^{-1}$ and $4p_{3/2}^{-1}$ manifolds, respectively, as the infrared pump intensity is varied. Dashed lines denote linear fits.
	}
	\label{fig:3}
\end{figure}

In the attosecond probe step, a circularly polarized XUV pulse couples the $4p^{-1}$ and $3d^{-1}$ manifolds. Within first-order perturbation theory, the resulting circular-dichroic transient-absorption signal reads \cite{Santra2011,Pabst2012A2,WuMX2016}
\begin{equation}
\Delta A(\omega,t_d)
=
\frac{4\pi\omega}{c}\,
\mathrm{Im}\sum_{ijk}\rho^{+}_{ij}(t_d)\,
\frac{d_{+}^{jk}d_{-}^{ki}-d_{-}^{jk}d_{+}^{ki}}
{E_{kj}-\omega-i\Gamma_k/2},
\label{eq:ana}
\end{equation}
where $\rho^{+}_{ij}(t_d)$ is the ionic density-matrix element evaluated at the arrival of the XUV probe, $E_{kj}\equiv E_k-E_j$, and $\Gamma_k$ is the linewidth of the core-excited state $k$. Here $d_{\pm}^{jk}=\langle j|\hat d_{\pm}|k\rangle$, where $\hat d_{\pm}$ are the dipole operators associated with circularly polarized probe interactions. The two dipole products encode the LCP and RCP Raman pathways $i\!\to\!k\!\to\!j$ (absorption followed by emission). To isolate the dichroic content of Eq.~(\ref{eq:ana}), we collect the helicity-dependent dipole products into a single effective operator that governs the CD response,
\begin{equation}
\hat{O}_{\rm CD}=\hat d_{+}\hat P_{e}\hat d_{-}-\hat d_{-}\hat P_{e}\hat d_{+},
\label{eq:ocd}
\end{equation}
where $\hat P_e=\sum_{J_e M_{J_e}}|\alpha J_e M_{J_e}\rangle\langle\alpha J_e M_{J_e}|$ projects onto the relevant excited-state manifold. Irreducible-tensor algebra shows that $\hat{O}_{\rm CD}$ reduces to the $q=0$ component of a rank-1 tensor, $\hat T^{(1)}_{0}=\sqrt{2}\,\bigl[\hat d\otimes(\hat P_{e}\hat d)\bigr]^{(1)}_{0}$, i.e., it transforms in the same way as $\hat{J}_z$. This is the same tensor structure that underlies the Thole-Carra X-ray magnetic CD sum rules \cite{Thole1992,Carra1993}. By the Wigner-Eckart theorem, the matrix elements of $\hat{O}_{\rm CD}$ are separately proportional to the corresponding matrix elements of $\hat L_z$ and $\hat S_z$ (see Supplementary Information). In the present scheme this operator acts on the time-evolving ionic density matrix instead of a static ground state, so that at each pump-probe delay $t_d$ the CD signal tracks the instantaneous $\hat L_z$ and $\hat S_z$ of the ionic vacancy. Because spin-orbit coupling intertwines the orbital and spin degrees of freedom, the CD signal maps linearly onto the ring-current dynamics and the accompanying spin dynamics. Therefore, attosecond CD transient absorption provides all-optical time-domain access to both the orbital current and the accompanying spin evolution. Since $\rho^{+}_{ij}(t_d)$ comprises stationary populations together with SO-driven coherences oscillating at $\Delta_{\rm SO}$, Eq.~(\ref{eq:ana}) separates naturally into a delay-independent dc background set by the populations and a delay-dependent ac modulation set by the coherences. The dc and ac components of the CD signal therefore inherit the static and alternating ring currents, respectively.

\subsection*{Retrieval and analysis}
To extract quantitative information on the ionic vacancy ring current from the computed attosecond transient-absorption spectra, we analyze the CD response obtained with co-rotating and counter-rotating attosecond probe helicities. As shown in Fig.~\ref{fig:3}(a) and (b), three dominant absorption features appear in the transient-absorption spectra, corresponding to the allowed transitions from the $4p_{J}^{-1}$ to the $3d_{J'}^{-1}$ core-excited manifolds.
Overall, co-rotating probing gives stronger absorption than counter-rotating probing. Importantly, one absorption line remains essentially delay-independent, whereas the other two exhibit pronounced delay-dependent oscillations. This behavior arises from quantum-path interference between transitions originating from different $4p^{-1}$ initial manifolds and converging on a common $3d^{-1}$ final manifold. The delay-dependent oscillations directly reflect the preserved coherence between the spin-orbit-split vacancy states.

From the absorption spectra, we construct the CD absorption difference between LCP and RCP probes, yielding the delay-dependent CD signal shown in Fig.~\ref{fig:3}(c). Averaging the CD spectrum over the delay gives its dc component, shown in Fig.~\ref{fig:3}(e). The distinct absorption peaks in this averaged spectrum encode the contributions from the spin-orbit-split vacancy manifolds. Figures~\ref{fig:3}(g) and~\ref{fig:3}(h) show that the state-resolved dc components depend linearly on the corresponding static ring-current magnitudes for the $4p_{1/2}^{-1}$ and $4p_{3/2}^{-1}$ manifolds as the infrared pump intensity is varied. This demonstrates that the state-resolved dc component of the CD transient-absorption signal provides a direct quantitative inversion of the ring-current magnitude and direction.
In addition to the dc contribution, the CD signal contains a delay-dependent ac component that captures the dynamical modulation of the system. As shown in Fig.~\ref{fig:3}(d), this ac component is obtained by subtracting the delay-averaged background from the CD signal and exhibits a periodic evolution from a Fano line shape to a Lorentzian line shape, which can be ascribed to the dipole phase~\cite{Pfeiffer12A,Ott2013}. This line-shape oscillation reflects how the spin-orbit coherence accumulates a phase $\Delta_{\rm SO}\,t_d$ that grows linearly with the pump-probe delay, shifting the relative phase between the resonant ionic dipole and the probe field and thereby cycling the CD profile between Lorentzian and Fano shapes. The CD signal can be approximated by $\Delta A(\omega,t_d)=\alpha(\omega)+\beta(\omega)\cos(\Delta_{\rm SO}t_d+\phi(\omega))$.

Integrating the ac signal over photon energy and comparing it with the corresponding ac component of the vacancy ring current [Fig.~\ref{fig:3}(f)] reveals excellent agreement in their temporal evolution. This correspondence shows that the ac component of the attosecond CD transient-absorption signal tracks the ring-current dynamics in real time, enabling an all-optical measurement of its instantaneous evolution with attosecond resolution.

\begin{figure}[!htbp]
	\centering
\includegraphics[width=0.78\linewidth]{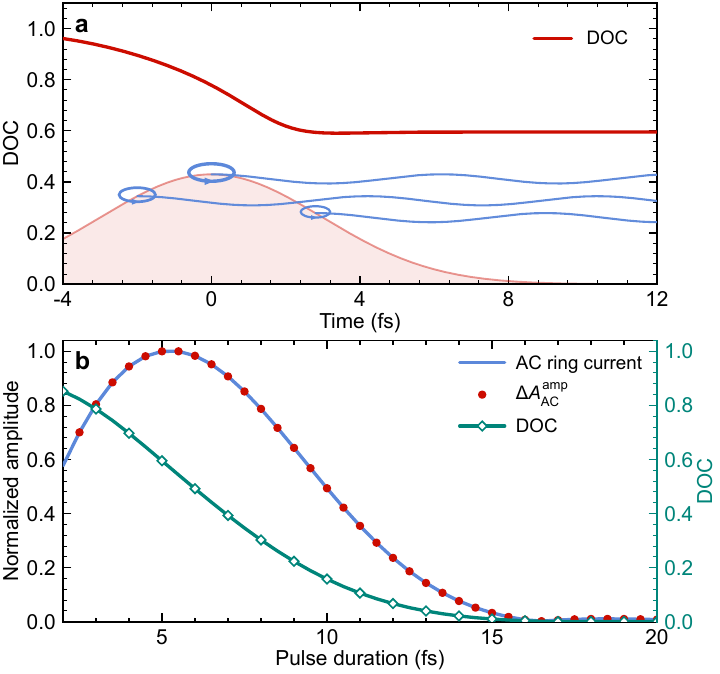}
	\caption{Dependence of the degree of coherence, ac ring current, and CD modulation on the pump-pulse duration. (a) Degree of coherence (DOC) between the $M_J=1/2$ components of the spin-orbit-split $4p_{1/2}^{-1}$ and $4p_{3/2}^{-1}$ vacancy states during the pump pulse. Ring-current contributions injected at different times within the LCP pump pulse acquire different spin-orbit phases, leading to partial cancellation when coherently summed over the entire pulse. The shaded profile indicates the pump-pulse envelope. (b) Normalized amplitudes of the ac ring current and the CD modulation, together with the DOC, as functions of the pump-pulse duration.}
	\label{fig:4}
\end{figure}

\subsection*{Pump-duration control of the coherence}
We now examine how the CD signal depends on the infrared pump-pulse duration.
Figure~\ref{fig:4}(a) shows the degree of coherence between the $M_J=1/2$
components of the spin-orbit-split vacancy states during the pump pulse.
As ionization proceeds, the DOC gradually decreases because ring-current
contributions injected at different ionization instants acquire different
spin-orbit phases and partially cancel when coherently summed over the pulse. Figure~\ref{fig:4}(b) shows that the ac ring current formed by the coherent superposition of the spin-orbit-split vacancy manifolds first increases and then decreases with increasing pump-pulse duration, reaching a maximum near $5\,\mathrm{fs}$. This time scale is set by the spin-orbit period $T_{\rm SO}=2\pi\hbar/\Delta_{\rm SO}\approx 6.2\,\mathrm{fs}$ in Kr$^+$. Optimal coherence is built up when the pump duration is comparable to $T_{\rm SO}$: short enough to limit dephasing yet long enough to inject sufficient coherence. This behavior reflects the competition between the build-up of the ac ring current induced by ionization injection and the accumulation of decoherence over the pump duration. The modulation depth extracted from the attosecond CD absorption spectra follows the same pump-duration dependence, as also shown in Fig.~\ref{fig:4}(b). This explains why only the ground-state ring current was observed in the measurement with a long pulse of 40~fs duration \cite{eckartUltrafastPreparationDetection2018b}, whereas the spin-orbit wave-packet evolution can be seen for a pulse duration of 7~fs \cite{Fleischer11_2,Fechner2014}. It thus establishes the pump-pulse duration as an effective control parameter for tuning the observability of ring-current dynamics via attosecond CD spectroscopy. We point out that, although the vacancy coherence is mainly discussed here, the ion-photoelectron entanglement is in fact reflected in the reduced density matrix of the ion.

\section*{Discussion}
We have theoretically established an all-optical attosecond circular-dichroic transient-absorption scheme for probing vacancy ring currents and their spin-orbit-entangled dynamics created by strong-field ionization in atoms. Combining analytic derivations with numerical simulations for krypton, we have shown that the delay-independent and delay-dependent components of the CD absorption signal provide a quantitative mapping onto the static and alternating ring currents. This mapping arises from the selective sensitivity of the circular-dichroic transition operator to the orbital angular-momentum projection along the probe direction and to the associated spin component, as formalized through the Wigner-Eckart theorem. Our analysis further demonstrates that the pump-pulse duration controls the degree of ionic coherence injected by strong-field ionization, defining the observable temporal range of the ring-current and spin dynamics through the competition between coherence build-up and decoherence. Because the proposed detection principle relies only on spin-orbit coupling and angular-momentum selection rules, it applies directly to atomic systems and may extend to molecular systems, in which vibrational, rotational and electronic degrees of freedom are intertwined. The present results establish attosecond circular-dichroic transient absorption as a powerful electronic-state-resolved time-domain analog of X-ray magnetic circular dichroism at the atomic scale.

\section*{Methods}
\subsection*{Ionization-coupling master equation}
The reduced density matrix $\rho^{+}$ of Kr$^+$ is propagated according to Eq.~(\ref{eq:one}). The basis comprises the fine-structure-resolved $4p^{-1}_{3/2}$, $4p^{-1}_{1/2}$ and $4s^{-1}_{1/2}$ valence-vacancy states and the $3d^{-1}_{5/2}$ and $3d^{-1}_{3/2}$ core-excited states. The atomic structure parameters are calculated using the Flexible Atomic Code~\cite{Gu2008FAC}. The level energies are corrected using available experimental data~\cite{Sugar1991,goulielmakisRealtimeObservationValence2010,wirthSynthesizedLightTransients2011c}. The level energies and reduced dipole matrix elements used in the calculation are listed in Supplementary Tables~1 and~2, respectively.

The ionization-injection term $S_{\rm ion}(t)$ uses the $m_l$-resolved nonadiabatic tunneling rates of Ref.~\cite{barthNonadiabaticTunnelingCircularly2011}. 
{
The ionization term is constructed in the ionic basis as
\begin{equation}
S_{\rm ion}(t)=\frac{\rho_N(t)}{2}
\sum_{M_L=-1}^{1}\sum_{M_S=\pm1/2}
w_{\eta}^{1,-M_L}(t)\,
|M_L,M_S\rangle\langle M_L,M_S|,
\label{eq:methods_injection}
\end{equation}
where $\rho_N(t)$ is the neutral population and $|M_L,M_S\rangle$ denotes a $4p^{-1}$ vacancy state with $L=1$ and $S=1/2$. The ionic and electron orbital projections satisfy $M_L=-m_l$, so $w_{\eta}^{1,-M_L}(t)$ is the ionization rate of the electron with $m_l=-M_L$ for pump helicity $\eta=+1$ (LCP) or $-1$ (RCP). The vacancy states are projected onto the spin--orbit-coupled basis $|J,M_J\rangle$. For $M_J=\pm1/2$, this produces coherence between $4p^{-1}_{3/2}$ and $4p^{-1}_{1/2}$, whereas $M_J=\pm3/2$ belongs only to $4p^{-1}_{3/2}$ (see Supplementary Information).
}
 The decay term $D_{\rm decay}(t)$ accounts for Auger decay of the $3d^{-1}$ states with a lifetime of $7.5$~fs~\cite{Jurvansuu2001}.

\subsection*{Laser parameters}
The pump is a circularly polarized pulse of central wavelength $800$ nm, full width at half maximum duration $5\,\mathrm{fs}$ (unless stated otherwise) and peak intensity $2\times10^{14}\,\mathrm{W/cm^2}$. The probe is an isolated attosecond XUV pulse of $150\,\mathrm{as}$ duration centered at $80\,\mathrm{eV}$ and peak intensity $1\times10^{10}\,\mathrm{W/cm^2}$, with left- or right-circular polarization, delayed by $t_d$ with respect to the pump. Both pulse durations refer to the intensity full width at half maximum. The fields have Gaussian envelopes, and their common propagation direction is chosen as the quantization $z$ axis.

\subsection*{Circular-dichroic absorption and ring current}

{
For probe helicity $\lambda$, the transient-absorption spectrum is computed from the Fourier transforms of the time-dependent ionic dipole $\mathbf d_{\lambda}(t,t_d)$ and the light field $\mathbf E_{\lambda}(t,t_d)$,
\begin{equation}
A_{\lambda}(\omega,t_d)=\frac{4\pi\omega}{c}\,
{\rm Im}\left[
\frac{\mathbf d_{\lambda}(\omega,t_d)\cdot\mathbf E_{\lambda}^{\ast}(\omega,t_d)}
{|\mathbf E_{\lambda}(\omega,t_d)|^2}
\right].
\label{eq:methods_absorption}
\end{equation}
The CD signal is defined as $\Delta A(\omega,t_d)=A_{+}(\omega,t_d)-A_{-}(\omega,t_d)$. Within first-order perturbation theory, Eq.~(\ref{eq:ana}) is obtained under the short-probe approximation after the pump pulse (see Supplementary Note~4). The delay-independent component is the average of $\Delta A(\omega,t_d)$ over the scanned delay range, and the ac component is obtained by subtracting this average from the delay-dependent signal.
}

 Within the $4p^{-1}$ manifold, the vacancy ring current is evaluated from $I_{\rm ring}(t)\propto{\rm Tr}[\rho^{+}(t)\hat L_z]$ (see Supplementary Note~3). The degree of coherence is defined as $|\rho^+_{ij}(t)|/\sqrt{\rho^+_{ii}(t)\rho^+_{jj}(t)}$ between the $M_J=1/2$ components of $4p^{-1}_{3/2}$ and $4p^{-1}_{1/2}$.

\subsection*{Data availability}
The data supporting the findings of this study are available within the article and its Supplementary Information. Source data are provided with this paper.

\subsection*{Code availability}
The custom codes used for the numerical simulations and data analysis in this study are available from the corresponding authors upon reasonable request.

\bibliography{refs}

\backmatter

\bmhead{Acknowledgements}
We acknowledge funding from the National Natural Science Foundation of China (Grants Nos. 12595341, 12234020, 12474281, 12450403, 12274461).

\bmhead{Author contributions}
Z.Z. and J.Z. conceived the project. G.B. performed the calculations with contributions from Z.L. and J.L. All authors analyzed the results and contributed to writing the manuscript.

\bmhead{Competing interests}
The authors declare no competing interests.

\end{document}